\pdfoutput=1
\documentclass{acm_proc_article-sp}
\usepackage{hyperref}
\begin{document}

\title{Personalized Web Search}

\numberofauthors{1} %  in this sample file, there are a *total*
\author{
% 1st. author
Li Zhou\\
       \affaddr{Language Technologies Institute}\\
       \email{lizhou@cs.cmu.edu}
}

\maketitle
\begin{abstract}
Personalization is important for search engines to improve user experience. Most of the existing work do pure feature engineering and extract a lot of session-style features and then train a ranking model. Here we proposed a novel way to model both long term and short term user behavior using Multi-armed bandit algorithm. Our algorithm can generalize session information across users well, and as an Explore-Exploit style algorithm, it can generalize to new urls and new users well. Experiments show that our algorithm can improve performance over the default ranking and outperforms several popular Multi-armed bandit algorithms.
\end{abstract}



 % NOT required for Proceedings

\section{Background of this IRLab}
\textbf{Note that this section is not a technical session, it just describes how I did this IRLab}. My IRLab project started from last semester. At the beginning, I wanted to do personalization on Web Search, because it is a hot topic in both academic and industry world. I also found a personalization dataset which was released as part of WSDM 2014 workshop. At first, I tried to use Hidden-semi markov model to model the short term user behavior, where each user's intent is a hidden state and the observations are queries within the same session period. I implemented this method during the summer. However it turned out that it did not work well: the average number of queries in each session period is about 3, so the data is really sparse; on the other hand, Hidden-semi markov model has a lot of parameters. As a result, the model I got overfit the data. I tried several techniques to fix this, such as decrease the number of parameters and keep only the sessions that contains more than 5 queries. However my results showed that it still did not work well. 

This semester my capstone project is to use Multi-armed bandit algorithm for news personalization. One advantage of Multi-armed bandit is that it can handle the `cold start' problem, which means it can learn to personalize a new user very quickly. Most of the academic paper apply Multi-armed bandit to news or ads recommendation, none of them have tried Multi-armed bandit to Web Search, so for this IRLab I decided to apply Multi-armed bandit to Web Search dataset. Web Search is a harder problem than news recommendation, and one big challenge is how to make use of the session information of users to model the long term and short term user behavior. This is a totally different problem from my capstone project because in my capstone project, I focused on using Multi-armed bandit algorithm on news personalization for new users; However in IRLab, I focused on Web Search personalization, and I proposed a new algorithm for modeling long term and short term user behavior using session information. \\
Code: \url{https://bitbucket.org/lizhou_cmu/personalization} \\
Site: \url{https://sites.google.com/site/personalizedwebsearch} \\

\section{Introduction}
The goal of search engine is to match users' intent with indexed documents. Traditional search engine treat a user's queries as his/her intent. However queries are not sufficient for expressing user intent. For example, if a user searches a movie name, e.g. `Forrest Gump', what does this user really want? does he want to watch this movie, or just want to read the reviews, or want to find out who is the star? We need more context information in order to infer his/her interests. If the previous query of this user is `movie trailer online', then we have a high confidence that this user wants to watch the trailer of `Forrest Gump'. If the search engine just return a list of results based on a user's query and ignore the user's specific interests and/or search context, then the identical query from different users or in different contexts will generate the same set of results displayed in the same way for all users, and users' intent will be hard to satisfy. 

Personalization is one way to solve this problem. Basically personalization techniques put a search in the context of the user's interests, and consider two cases \cite{bennett2012modeling}: the long term user behavior in the past and the short term user behavior in the current session. These two types of contextual information are then used to infer the user intent. For a given query, a personalized search provide different results for
different users based on the current user's intent.

Search session is a series of intent-related user's requests to the search engine\cite{ustinovskiy2013personalization}. Studies\cite{cao2008context} \cite{shen2005context} has been shown that a user tends to refine the queries or explore related information about his or her search intent in a session. There are about five types of refinement, a) spelling correction: refine `MSN messnger' to `MSN messenger'; b) peer queries: refine `SMTP' to `POP3'; c) Acronym: refine `BAMC' to `Brooke Army Medical Center'; d) Generalization: refine `Washington mutual home loans' to `home loans'; e) Specialization: refine `Nokia N73' to `free themes Nokia N73'.

Long term behavior of a user is the user profile built based on the history behavior of that user. Most of the current work\cite{matthijs2011personalizing} \cite{tan2006mining} make use of the click data and try to infer user's interests. For example if the user tends to click technology articles, then next time when he search `Apple', the search engine will show `Apple' the company rather than `apple' the fruit.

In this paper, we tried two methods to model the short term user behavior within one session. The first method is Hidden-semi markov model, where we assume the user intent is a hidden variable, and each user intent will generate a sequence of queries. After the last queries satisfy the user's needs, or the user gives up trying any other queries under such intent, then the hidden intent transit to another hidden variable and again generate a sequence of queries. The second method we tried is to use Multi-armed bandit techniques. We group different sessions into a few clusters. And while interact the user within one session, we try to identify which cluster the current session is in, and hence infer the user's intent. We also propose a long term behavior modeling algorithm. We build the user interests profile in a online style and adjust this profile based on user's interact with our algorithm.

The remainder of this paper is structured as follows, in section 3 we will talk about the existing methods for solving personalization problems, and then we describe a short term behavior modeling algorithm which is finally abandoned by us. In section 5 we will talk about our latest proposed algorithm and how we model long term and short term behavior. Finally we will show our results on the Yandex web search dataset\cite{yandexdata}.
\section{Related Work}
Personalization is a very popular topic. There are have been quite a few workshops on personalization\cite{yandexdata}\cite{nipsworkshop}. Most work focuses on the following problems: a) when should we do personalization and shouldn't? b) how to model long term user behavior? c) how to model short term user behavior?
\subsection{To Personalize or not to Personalize}
Personalization does not guaranteed to give us good results. \cite{teevan2010potential} and \cite{teevan2008personalize} showed that there is a lot of variation across queries in the benefits that can be achieved through personalization. For some queries, everyone who issues the query is looking for the same thing. For other queries, different people want very different results even though they express their need in the same way. They characterize queries using a variety of features of the query, the results returned for the query, and people's interaction history with the query. Using these features they build predictive models to identify queries that can benefit from personalization. Experiments suggested that for popular queries or queries with very similar click behavior between different people, no personalization strategy should be adopted. Similarly, \cite{dou2009evaluating} defined the click entropy to measure variation in information needs of users under a query. And experimental results showed that personalized Web search yields significant improvements over generic Web search for queries with a high click entropy. For the queries with a low click entropy, personalization methods performed similarly or even worse than generic search. As personalized search had different effectiveness for different kinds of queries, they argued that queries should not be handled in the same manner with regard to personalization, and the click entropy can be used as a simple measurement on whether a query should be personalized. 
\subsection{Long Term User Behavior Modeling}
Long term behavior modeling focuses on how to build a profile for each user based on history data, and then re-rank search results according to user profile. The key problem is how to represent and learn each user profile. Typically there are two ways, the first is to represent user profile using history queries, clicked urls, and many other features. The second method is to map user interests into a set of topics. \cite{matthijs2011personalizing} build a Firefox add-on and capture features such as unigram, noun phrase etc. Each term is scored by TF, TF-IDF and BM25. These terms and scores are treated as the user profile. Then they use simple score function such as inner product and language model to calculate the similarity of the user profile and the url snippet, and further re-rank the search results. \cite{sontag2012probabilistic} build a language model for each user as the user profile. The language model is built based on the queries and the query topics of that user. Then they train a probabilistic models to re-rank the default search results. \cite{teevan2011understanding} analyze the query logs by looking at navigation behavior across all users. These navigation behavior are used to construct the user profile. \cite{tan2006mining} explored how to exploit long-term search history, which consists of past queries, result documents and clickthrough, as useful search context that can improve retrieval performance. They estimate a query language model which is a weighted average of the unit history models of past queries, and they use EM algorithm to estimate the model.

There are also a lot of works map user interests into a set of topics. \cite{ma2007interest} map user interests to categories in the ODP taxonomy. Then they uses URLs organized under those categories as training examples to classify search results into various user interests at query time. \cite{chirita2005using} do re-rank based on the distance between a user profile defined using ODP topics and the sets of ODP topics covered by each URL returned in regular web search. The distance metrics they defined is very complex, it contains word semantic distance based on WordNet and the depth of the categories. \cite{bennett2010classification} \cite{eickhoff2013personalizing} classify urls into 1 of the 219 topical categories from the top two levels of the ODP. \cite{qiu2006automatic} learn a user topic perference vector based on history data, and then adopted topic-sensitive PageRank.

\subsection{Short Term User Behavior Modeling} 
Short term behavior modeling focuses on infer current user's intent based on behaviors in the current session. \cite{cao2008context} proposed a context-aware query suggestion approach which is in two steps: in the offline step, queries are clustered by a click-through bipartite, and a concept sequence suffix tree is constructed as the query suggestion model; in the online step, a user's search context is captured by mapping the query sequence submitted by the user to a sequence suffix tree. \cite{white2010predicting} investigate the effectiveness of activity-based context. Features include queries, SERP clicks, post-SERP navigation. They assign ODP category labels to URLs and call it query model, they also build context model based on actions that occur prior to the current query in the search session. Finally they build intent model which is the weighted average of query model and context model. Use all these model they get the topics of current session and then do re-ranking. \cite{ustinovskiy2013personalization} \cite{xiang2010context} defined a set of features such as the cosine and jaccard distance between the search results and the query. And then they trained a learning to rank model. \cite{liao2013vlhmm} adopted vlHMM to model session behavior, where each query in the session has a latent intention conditioned on the former n query, also states in the vlHMM are represented as feature vectors. \cite{white2013enhancing} mine historic search-engine logs to find other users performing similar tasks to the current user and leverage their on-task behavior to identify web pages to promote in the current ranking. Features include query, clicked urls and webpage content. \cite{kiseleva2013discovering} decomposed web sessions into non-overlapping segments and learned the temporal context for each segment. The goal is to discover temporal hidden contexts in the web search sessions. 

Finally, \cite{bennett2012modeling} and \cite{bennett2012modeling} make use of both long term and short term context to re-rank search results. Main feature they use are previously clicked urls.

\subsection{Existing Multi-armed Bandit Approach to Personalization}
Multi-armed bandit has been applied to news personalization. News personalization is quite different from Web Search. In news personalization, we assume that there are a set of candidate articles, and each time we only select one of them to the user, and after get feedback from the user, we udpate our model. The feedbacks are click/not click in our case. Here we talk about two algorithms, which will be our baseline algorithm. 
\subsubsection{LinUCB\cite{li2010contextual}}
LinUCB try to minimize the regret which is defined as the expectation of reward of best article and the expectation of reward of article selected by the algorithm:
\begin{align}
R(T) = E[\sum_{i=1}^T r_{t,a*}] - E[\sum_{i=1}^T r_{t, a_t}]
\end{align}
The reward in our case is the click-through rate. LinUCB model the expected reward of an article as a linear function of the context's feature vector
\begin{align}
E[r_{t,a}|x_{t,a}] = x^T_{t, a} \theta^*_a
\end{align}
So to predict the reward of an article, it is now a linear regression problem. However unlike traditional linear regression problem, we also have a confidence interval associated with each of our predicted value, that is, the estimated reward of an article is
\begin{align}
a_t = \arg\max(x^T_{t,a}\theta_a + \alpha \sqrt{x^T_{t,a}A_a^{-1}x_{t,a}})
\end{align}
where $A_a = D_a^TD_a + I_d$
The last part of the above equation is the confidence interval. What LinUCB does is to always select the article which has the highest upper confidence interval bound.

\subsubsection{Thompson Sampling}
Thompson sampling \cite{chapelle2011empirical} is a very old heuristic algorithm yet recently received a lot of attention. Thompson sampling is a Bayesian style algorithm, it model the reward likelihood as a parameteric function $P(r|a,x,\theta)$, where $\theta$ is the parameter vector for the model. Given some prior distribution $P(\theta)$ on these parameters, the posterior distribution of these parameters is given by the Bayes rule, $P(\theta|D) \propto \prod P(r_i|a_i, x_i, \theta)P(\theta)$. Normally the reward is a stochastic function of the action, context and then unknown true parameter $\theta^*$, and we want to choose the article that maximizeing the expected reward $\max_a E(r|a, x, \theta^*)$. However we do not know the true parameter $\theta^*$, so we are actually choosing the article based on the following probability with each article
\begin{align}
\int I(E(r|a, x, \theta) = \max_{a'}E(r|a', x, theta))P(\theta|D) d\theta
\end{align}
Where $I$ is the indicator function. Bascially we are selecting an article according to its probability of being optimal. When implementation, what we do this simple: sample a $\theta$ from current posterior distribution, calculate the probability of each article being optimal according to the above equation, and then select an article and receive reward from users (clicked or not clicked), finally we update the posterior distribution based on the rewward. Normally the prior and posterior distribution are all Gaussian distribution \cite{agrawal2012thompson}.
\section{Abandoned Proposed Method}
This is the first algorithm I came up with for short term personalization. Search sessions record the user behavior within a short period. Session segmentation normally based on time span or similarities between queries \cite{gayo2009survey}. Within a session, we assume that there are at last one and maybe more hidden intent. For example, during one session, a user may want to trade-in his `IPhone 3GS', so he first search `IPhone trade-in', and then search `IPhone trade-in Bestbuy', and then search `IPhone 3GS trade-in Bestbuy price'. Another example is that if a user want to have some nice dinner with his friends in Mount Washington, Pittsburgh, he may first search `restaurant mount washington', and then search `seafood mount washington pittsburgh' and then maybe `lobster mount washington pittsburgh' or `seafood buffet mount washington pittsburgh'. As we can see, there is a intent behind each of the sequence of the queries, and people are refining their queries to express their intent. Hidden-semi markov model\cite{yu2010hidden} is a perfect model for modeling such sequential transition. In the Hidden-semi markov model, there is a hidden variable which genearate a sequence of observations. The difference between it and the hidden markov model is that in hidden markov model, a hidden variance only generate one observation. A descriptive structure in figure 1 showed how we model session behavior.\\
%\hspace*{-1.0in}
\begin{figure*}
\centering
\includegraphics[scale=1]{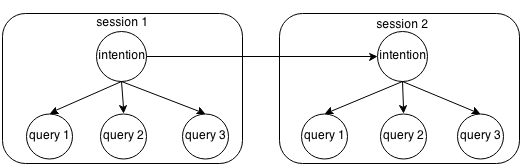}
\label{fig:hsmmgraph}
\caption{Modeling session behavior as Hidden-semi markov model}
\end{figure*}

Assume a discrete-time Markov chain with the set of (hidden) states $S=\{1, ..., M\}$. The state sequence is denoted by $S_{1:T} = S_1, ..., S_T$, where $S_t \in S$ is the state at time t. A realization of $S_{1:T}$ is denoted as $s_{1:T}$, and $S_{[t]} = i$ means state i starting and ending at $t$ with duration 1, and $S_[t_1:t_2] = i$ means state i starting from $t_1$ and ending at $t_2$. Denote the observation sequence by $O_{1:T} = O_1, ..., O_T$, where $O_t$ is the observation at time $t$. Then we can define the state transition probablity from $(i, d')$ to $(j, d)$ by
\begin{align}
a_{(i, d')(j, d)} = P(S_{[t+1:t+d]}=j|S_{[t-d'+1:t]}=i)
\end{align}
which is the probability of stay in state $i$ for duration $d'$ and then transite to state $j$ and stay for duration $d$. We can also define the emission probability by
\begin{align}
b_{j, d}(o_{t+1:t+d}) = P(o_{t+1:t+d}|S_{t+1:t+d}=j)
\end{align}
which is the probability of emit observation $o_{t+1:t+d}$ within time span $t+1:t+d$. The initial distribution of the state is defined as 
\begin{align}
\pi_{j, d} = P[S_{t-d+1:t}=j]
\end{align}
\subsection{Inference}
Similar to Hidden markov model, we use forward-backward algorithm to infer the MLE of state sequence. The forward variables for HSMM is defined by
\begin{align}
\alpha_t(j, d) = P(S_{t-d+t:t} = j, o_{1:t}|\lambda)
\end{align}
and the backward variables for HSMM is defined by
\begin{align}
\beta_(t,d) = P(O_{t+1:T}|S_{[t-d+1:t]} = j, \lambda)
\end{align}
so the forward-backward algorithm for HSMM is \cite{yu2010hidden}\cite{rabiner1989tutorial}:
\begin{align}
\alpha_t(j,d) = \sum_{i \in S \backslash \{j\}} \sum_{d'\in D} \alpha_{t-d}(i, d') a_{(i,d')(j,d)} b_{j, d}(o_{t-d+1}:t) \\
\beta_t(j, d) = \sum_{i \in S \backslash \{j\}} a_{(j, d)(i, d')} b_{i, d'}(O_{t+1:t+d'}) \beta_{t+d'}(i, d')
\end{align}
After the forward variables and backward variables are determined, we can calculate the probability that state $j$ start at $t-d+1$ and ends at $t$, with duration $d$, given partial observed sequence $o_{1:t}$
\begin{align}
P(S_{t-d+1:t} = j | o_{1:t}, \lambda) = \frac{\alpha_t(j, d)}{\sum_{j,d}\alpha_t(i, d')}
\end{align}
and the predicted probability that state j will start at $t+1$ and end at $t+d$ with duration d, given partial observed sequence $o_{1:t}$ by
\begin{align}
P(S_{t+1:t+d}=j|o_{1:t}, \lambda) = \frac{\sum_{i\neq j,d'}\alpha_t(i, d')a_{(i,d')(j,d)}}{\sum_{i,d'}\alpha_t(i, d')}
\end{align}
The posterior probabilites for given entire observation sequence $o_1:t$ can be determined by
\begin{align}
\eta_t(j, d) &= P(S_{t-d+1, t}=j, o_{1:T}|\lambda) = \alpha_t(j, d)\beta_t(j, d) \\
\epsilon_t(i, j) &= \sum_{d'\in D} \sum_{d\in D}\epsilon(i, d'; j, d)
\end{align}
\subsection{Estimation}
We can re-estimate using the expectations:
\begin{align}
\hat{a}_{(i,d')(j, d)} &= \sum_t \epsilon_t(i, d'; j, d)/\sum_{j\neq i, d}\sum_t \epsilon(i, d'; j, d) \\
\hat{b}_{j, d}(o_1:o_d) &= \sum_t[\eta_t(j, d) I(o_{t+1:t+d} = O_1:d)] / \sum_t \eta_t(j,d)
\end{align}
To reduce the number of parameters, we used queries' cluster rather than queries' unigram as the observation variables. We trained a ranking model for each query cluster offline. And then predicting, we use HSMM model to predict the query cluster of the next query, and then use the corresponding ranking model to re-rank the search results.
\subsection{Why Abandon This Method}
I implemented this method during the summer. Based on an opensource project\cite{sm2} from JHU, I wrote my own HSMM model code. However I found out that I am too optimistic about this model. First, in the dataset, most of the sessions only have one or two queries, so it is not sufficient to estimate the HSMM model. Then I kept only the sessions that contain more than 5 queries, which slightly solved this problem.

Second, it is hard to estimate the stop condition of HSMM. General HSMM application such as those in speech or NLP do not have that much search space, so with sufficient data we can infer the probability of stop current sequence and transit to another hidden variable after staying in state $i$ for time $t$. However in our case users may stop because they are satisfied about the results, or because they just give up and leave. There are many factors that affect the stop condition, and based on my observation of the final model, the HSMM model I got is very random, and do not capture stopping behavior.

Third, HSMM is not a non-parametric model, so we have to set the total number of different hidden `intent' we have in the model. Moreover, since we want to avoid overfitting, we use query cluster instead of query word as obervations, so we have to set the total number of query clusters. Our results are sensitive to the two parameters.

Fourth, a query in our model is not depended on the previous query. However it should be. Users often refine their query based on their previous query, not just based on their intent. However if we make each query depend on the previous one, it would make the model even harder to infer and estimate, because the number of parameters to be estimated will again increase and more data is needed.

As a result, HSMM is a nice model in NLP area, but it is not that easy to apply to user session behavior model.

\section{Proposed Method}
I proposed to use Multi-armed bandit algorithm on Web Search for personalization. Since Multi-armed bandit select one article at a time, so we focus on the rank 1 results in the search result list, and try to improve the click-through rate of the rank 1 article.
\subsection{Short term behavior modeling}
Normally each session contains one intent such as `food in recent area' or `local weather'. Sessions with the same intent will be repeated across different users. For example, when people want to find a place to trade-in their old computer, they may all search queries such as `trade-in bestbuy' or `trade-in computer' etc. So sessions with this intent will be repeated again and again. Thus we can group sessions based on the queries, and then treat these sessions as our prior knowledge. When predicting online, we need to infer which session group this session is in, this can be done by a Generalized Thompson sampling \cite{li2013generalized}. Algorithm has the following step
\subsubsection{Step 1}
In the training data, for each session, we find out the urls been clicked. Then we build a hash table: the key is the clicked url, and the value is the query term. So for each clicked url, we have a set of associated query term. Then each session is represented by the query terms that is associated with all the clicked urls in that session.
\subsubsection{Step 2}
Based on Step 1, we get a set of sessions, and each session is associated with a set of query terms. We then run a LDA on this dataset, which each session is treated as a document. 
\subsubsection{Step 3}
For each topic we get from LDA, we collect all the sessions in that topic, and then train a RankSVM model for that topic.
\subsubsection{Step 4 (online)}
When predicting, we need to infer which session cluster the current session belongs to. To do that, we adopt Generalized Thompson sampling \cite{li2013generalized}. More specifically, we treat each pre-trained RankSVM model as the prior model $\{\varepsilon_1, \varepsilon_2, ..., \varepsilon_N\}$, then within the original 10 ten results, we reassign an article to rank 1 based on the following probability:
\begin{align}
P(a) = (1-\gamma)\sum_{i=1}^N \frac{w_{i,t}I(\varepsilon_i(x_t) = a)}{W_t} + \frac{\gamma}{K}
\end{align}
Here $\gamma$ is a smoothing parameter, and $w_{i, t}$ is the weight of prior model $i$ at time $t$. The idea behind this is that each model will vote a url of which they believe should be ranked at top 1, then we take the weighted average of their vote and do normalization. After we re-rank, we will recevie feedback from the user, that is whether the user clicked or not clicked the rank 1 url. Based on the feedback, we update the weight of each RankSVM model based on
\begin{align}
\forall_i, w_{i, t+1} &\leftarrow w_{i, t} \cdot \exp(-\eta \ell(f_i(x_t, a_t), r_t)) \\
W_{t+1} &\leftarrow ||w_{t+1}||_1
\end{align}
Here $\ell$ is the loss function, in our case it could be logarithmic loss: $\ell(\hat{r}, r) = I(r=1)ln1/\hat{r} + I(r=0)ln1/(1-\hat{r})$, $\eta$ is the learning rate and $f_i(x_t, a_t)$ is the prediction function provided by RankSVM. The idea behaind this is that we penalize when the model predict wrong with a loss function and adjust the weight of each RankSVM model after we observe users' behaviors. 
\subsubsection{Step 5}
In a real search engine system, after a pre-defined time period, e.g. 1 day, we can re-train these RankSVM models use data up to date. In this case we will update our model based on the latest user behavior.

One advantage of this algorithm is that Generalized Thompson Sampling converges very fast, so it can very quickly identify which session cluster the current session is in, that is, it can find a reasonable weight vector $w_t$ very fast.

\subsection{Long term behavior modeling}
We use thompson sampling with linear payoff \cite{agrawal2012thompson} to model long term user behavior. Again, we are focusing on the top 1 ranking of the search list. We still assume that there are 10 default search results, and we try to select one that we think the user will click with the highest probability and put it in the rank 1 position. The probability of being clicked by a user is modeled as a linear regression with a Gaussian error. Suppose the likelihood of reward $r_i(t)$ at time $t$ given context $b_i(t)$ and parameter $u$ were given by the pdf of Gaussian distribution $N(b_i(t)^T\mu, v^2)$, and Let
\begin{align}
B(t) &= I_d + \sum_{\tau=1}^{t-1} b_{a(\tau)}(\tau) b_{a(\tau)} (\tau)^T \\
\hat{\mu}(t) &= B(t)^{-1} (\sum_{\tau=1}^{t-1}b_{a(\tau)}(\tau)r_{a(\tau)}(\tau))
\end{align}
Then if the prior for $\mu$ at time $t$ is given by $N(\hat{\mu}(t), v^2B(t)^{-1})$, they posterior distribution at time $t+1$ is then
\begin{align}
P(\hat{\mu}|r_i(t)) & \propto P(r_i(t)\|\hat{\mu})P(\hat{\mu}) \\
& \sim N(\hat{\mu}(t+1), v^2B(t+1)^{-1})
\end{align}
Then the algorithm consist of the following steps:
\subsubsection{Step 1}
Sample $\hat{\mu}(t)$ from distribution $N(\hat{\mu}, v^2B^{-1})$
\subsubsection{Step 2}
Select article
\begin{align}
a_i(t) = \arg\max_i b_i(t)^T\hat{\mu}(t)
\end{align}
and observe reward (clicked or not clicked) $r_t$.
\subsubsection{Step 3}
Update 
\begin{align}
B &= B + b_{a(t)}(t)^T \\
f &= f+b_{a(t)}(t)r_t \\
\hat{\mu} &= B^{-1}f
\end{align}

The idea behind this is that we predict the click-through rate as a linear regression problem, and assume a Gaussian prior. Each time we observe the feedback from a user, we will update the posterior, and then sample a new parameter from the posterior distribution. Compare to a pure linear regression model or a ranking model, the advantage of this model is that as a Explore-Exploit algorithm, it can quickly learn the performance of a new article, while standard linear regression suffers overfitting when data is insufficient.

\section{Dataset}
We use the data provided by Yandex as part of WSDM 2014 Log-based Personalization workshop\cite{yandexdata}. The dataset includes user sessions extracted from Yandex logs, with user ids, queries, query terms, URLs, their domains, URL rankings and clicks. The user data is fully anonymized. Only meaningless numeric IDs of users are released. Some characteristics of the dataset:
\begin{enumerate}
\item Unique queries: 21,073,569
\item Unique urls: 703,484,26
\item Unique users: 5,736,333
\item Training sessions: 34,573,630
\item Test sessions: 797,867
\item Clicks in the training data: 64,693,054
\item Total records in the log: 167,413,039
\end{enumerate}
The time span for the training dataset is 27 days, and the time span for the test dataset is 3 days. However we do not have the ground truth of the test data, so we split training data into 24 days of training dataset and 3 days of test dataset.

In the dataset, each session contains one or more queries, and each queries contains 10 default search results, and which url the user clicked. All the behavior is associated with a time units. For example, the click log contains information about how many time units passed before the user clicked that url. Each log instance is either session metadata, or query action, or click action. Session metadata contains session Id, day id, and user id; query action contains session id, time passed, search id, query id, list of terms and list of url and domains; click action contains session id, time passed, search id and url id.

\section{Experiment Setting}
\subsection{Metric}
URLs are labeled using 3 grades of relevance: 0 (irrelevant), 1 (relevant), 2 (highly relevant). The labeling is based on dwell-time: 0 (irrelevant) grade corresponds to documents with no clicks and clicks with dwell time strictly less than 50 time units; 1 (relevant grade) corresponds to documents with clicks and dwell time between 50 and 399 time units (inclusively); 2 (highly relevant) grade corresponds to the documents with clicks and dwell time not shorter than 400 time units. In addition, the relevance grade of 2 assigned to the documents associated with the clicks which are the last actions in the corresponding sessions. Dwell time is the time passed between the click on the document and the next click or the next query. It is well-known that dwell time is well correlated with the probability of the user to satisfy her information need with the clicked document \cite{collins2011personalizing}.

Since we are using Multi-armed bandit algorithm, we are interested in only the click through rate of the rank 1 position of the search results. So our performance metric is the CTR of rank 1 position of the search list instead of the NDCG or MAP of the entire search list. 

\subsection{Feature construction}
I use the following 18 features described in Table 1. Note that our goal is not feature engineering, so I didn't spend huge amount of time to come up with new features. These features try to capture the url related click statistics under one session, one user and all the users. These features are used by RankSVM models for both short term and long term modeling.
\begin{table*}[!htbp]
\label{featuretable}
\caption{Features used by both long term and short term modeling}
\begin{tabular}{|l|l|l|}
\hline
fea id & fea name                     & fea description                                                                             \\ \hline
1      & short term relevance level 2 & in this session, the number of times that this url is level 2 relevent                      \\ \hline
2      & short term relevance level 1 & in this session, the number of times that this url is level 1 relevent                      \\ \hline
3      & short term relevance level 0 & in this session, the number of times that this url is level 0 relevent                      \\ \hline
4      & short term show              & in this session, the number of times this url is shown                                      \\ \hline
5      & short term missed            & in this session, the number of times this url is missed                                     \\ \hline
6      & short term skipped           & in this session, the number of times this url is skipped                                    \\ \hline
7      & long term relevance level 2  & in all the history data of this user, the number of times that this url is level 2 relevent \\ \hline
8      & long term relevance level 1  & in all the history data of this user, the number of times that this url is level 1 relevent \\ \hline
9      & long term relevance level 0  & in all the history data of this user, the number of times that this url is level 0 relevent \\ \hline
10     & long term show               & in all the history data of this user, the number of times this url is shown                 \\ \hline
11     & long term missed             & in all the history data of this user, the number of times this url is missed                \\ \hline
12     & long term skipped            & in all the history data of this user, the number of times this url is skipped               \\ \hline
13     & cross user relevance level 2 & in all the history data of all users, the number of times that this url is level 2 relevent \\ \hline
14     & cross user relevance level 1 & in all the history data of all users, the number of times that this url is level 1 relevent \\ \hline
15     & cross user relevance level 0 & in all the history data of all users, the number of times that this url is level 0 relevent \\ \hline
16     & cross user show              & in all the history data of all users, the number of times this url is shown                 \\ \hline
17     & cross user missed            & in all the history data of all users, the number of times this url is missed                \\ \hline
18     & cross user skipped           & in all the history data of all users, the number of times this url is skipped               \\ \hline
\end{tabular}
\end{table*}
\subsection{Compared Algorithms}
Note that our algorithm is a type of Explore-Exploit algorithm, so it can be used on top of any supervised learning algorithm such as RankSVM, LambdaMART, or RankNet. So here we focus on compare our algorithm with other Explore-Exploit algorithm:
\subsubsection{Default ranking}
This is the default ranking provided by Yandex, which is a very high baseline since it is a commercial search engine. The NDCG of default ranking is 0.794379.
\subsubsection{LinUCB \cite{li2010contextual}}
This is one of the most popular UCB style algorithm and have been treated as baseline for many papers. This algorithm has been described in the background section. Each time the algorithm select the articles that has the highest upper confidence bound.
\subsubsection{Thompson Sampling\cite{chapelle2011empirical}}
This is an algorithm that has been used by Yahoo! news personalization. It is a very efficient algorithm and has been proved to achieve lower regret than LinUCB.
\subsubsection{Our Proposed Algorithm}
The long term (based on Thompson sampling) and short term (based on Generalized thompson sampling) user behavior modeling algorithm proposed by us.

\section{Experiment Results}
\subsection{Results on Short term session data}
For short term session data, we first do cluster on all the sessions in training data, and then use Generalized Thompson sampling to infer which session group the current session is in, and then apply the corresponding ranking model. The result is shown is figure 2. 
\begin{figure}[!htbp]
\centering
\includegraphics[scale=0.37]{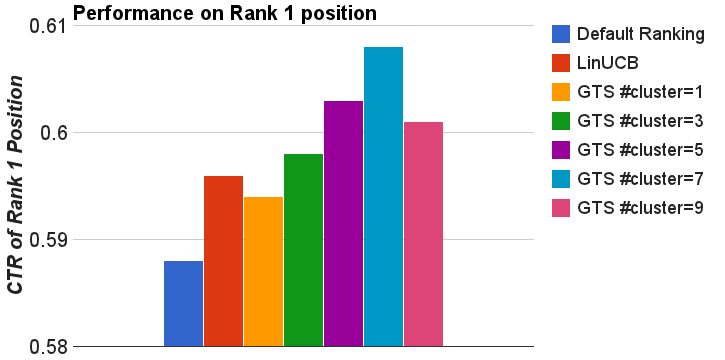}
\label{fig:res1}
\caption{Results on short term session data}
\end{figure}
From figure 2 we can see that when the number of cluster is 7, Generalized Thompson sampling achieved the best CTR on rank 1 position, it is 3.4\% higher than default ranking and 2.01\% higher than LinUCB. We can also see that when the number of cluster is 1, Generalized thompson sampling is still better than the default ranking, which means the RankSVM model itself can help improve the CTR of Rank 1 position.
\begin{figure}[!htbp]
\centering
\includegraphics[scale=0.45]{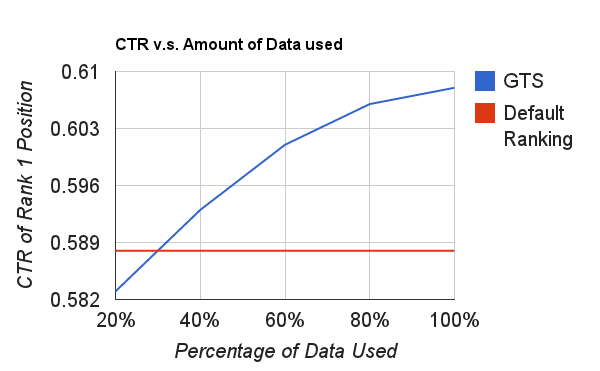}
\label{fig:res2}
\caption{Results on short term session data with different percentage of training data being used.}
\end{figure}
Figure 3 shows the performance of Generalized thompson sampling when different percentage of training data is being used. Since in our proposed algorithm, we have to do session clustering and train a ranking model on each of the cluster, so as we have more and more data, we can have higher and higher CTR.

\subsection{Results on Long term session data}
For long term user behavior modeling, our proposed model use Thompson sampling with linear payoffs, And the advantage is that it can explore the CTR of urls in long tail, while traditional supervised learning algorithm suffer data insufficient for long tail urls. First we show that our algorithm indeed sensitive to the learning rate, which is a common problem for all online learning algorithms. 
\begin{figure}[!htbp]
\centering
\includegraphics[scale=0.45]{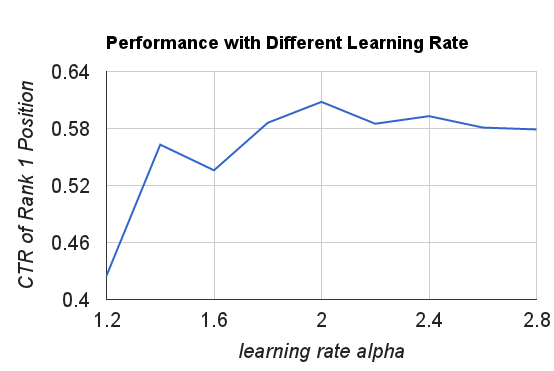}
\label{fig:res3}
\caption{Performance of Thompson sampling with linear payoffs with different learning rate}
\end{figure}
From Figure 4 we can see that the algorithm achieve highest CTR when the learning rate $\alpha=2$.
\begin{figure}[!htbp]
\centering
\includegraphics[scale=0.45]{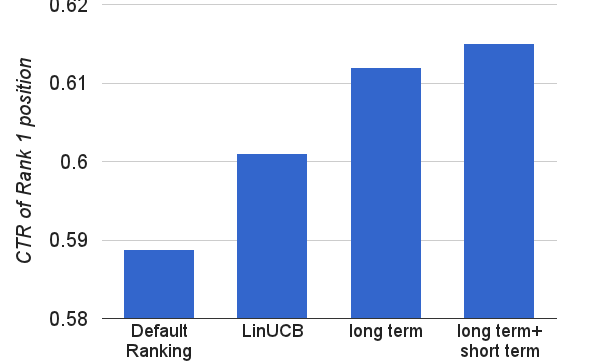}
\label{fig:res4}
\caption{Comparison of Default ranking and LinUCB with our proposed Long term and short term modeling algorithm}
\end{figure}
From Figure 5 we can see that our proposed long term and short term modeling algorithm can improve the CTR of rank 1 position of default ranking. Also note that the long term modeling can improve the CTR with 4.08\%, which is higher than the short term modeling. So we can see that long term modeling is more efficient than short term modeling. Also if we combine the long term and short term modeling, we can improve the default CTR on rank 1 position with 4.59\%.

\section{Conclusion}
We proposed a novel way to model both long term and short term user behavior using Multi-armed bandit algorithm. For short term, we first do clustering on session data and then use Generalized thompson sampling to identify which session cluster the current session is in, for long term, we use thompson sampling with linear payoff.  Experiments show that our algorithm can efficiently improve performance over the default ranking and outperforms LinUCB, a popular Multi-armed bandit algorithm.

\bibliographystyle{abbrv}
\bibliography{sigproc}  % sigproc.bib is the name of the Bibliography in this case
% ACM needs 'a single self-contained file'!
%\balancecolumns
%\balancecolumns
% That's all folks!
\end{document}